 \documentclass[preprint2]{aastex}
 \usepackage{epsfig}

\newcommand{\etal}{\textit{et al.}\ }

\shorttitle{Roe et al.}
\shortauthors{Altitude of Neptune's IR Bright Features}

\begin{document}


\title{The Altitude of an 
Infrared Bright Cloud Feature on Neptune from Near-Infrared
Spectroscopy\altaffilmark{1}}


\author{Henry G.\ Roe,\altaffilmark{2} 
James R.\ Graham,\altaffilmark{2} 
Ian S.\ McLean,\altaffilmark{3} 
Imke de Pater,\altaffilmark{2} 
E.E.\ Becklin,\altaffilmark{3} 
Donald F.\ Figer,\altaffilmark{4,5} 
Andrea M.\ Gilbert,\altaffilmark{2} 
James E.\ Larkin,\altaffilmark{3} 
N.A.\ Levenson,\altaffilmark{5} 
Harry I.\ Teplitz,\altaffilmark{6,7} 
and 
Mavourneen K.\ Wilcox\altaffilmark{3}  
}


\altaffiltext{1}{Data presented herein were obtained
at the W.M.\ Keck Observatory, which is operated as a scientific 
partnership among the California Institute of Technology, the 
University of California, and the National Aeronautics and Space
Administration.  The Observatory was made possible by the generous
financial support of the W.M.\ Keck Foundation.}
\altaffiltext{2}{Department of Astronomy, 601 Campbell Hall, University
of California, Berkeley, CA 94720-3411.}
\altaffiltext{3}{Department of Physics and Astronomy, UCLA, Los Angeles,
CA 90095-1562.}
\altaffiltext{4}{Space Telescope Science Institute, 3700 San Martin Drive,
Baltimore, MD 21218.}
\altaffiltext{5}{Department of Physics and Astronomy, Johns Hopkins
University, Baltimore, MD 21218.}
\altaffiltext{6}{Laboratory for Astronomy and Solar Physics, Code 681, 
Goddard Space Flight Center, Greenbelt, MD 20771.}
\altaffiltext{7}{NOAO Research Associate.}


\begin{abstract}

We present 2.03-2.30~$\mu$m near-infrared spectroscopy of Neptune
taken 1999 June 2 (UT) with the W.M.\ Keck Observatory's near-infrared
spectrometer (NIRSPEC) during the commissioning of the instrument.  
The spectrum is dominated by a bright cloud
feature, possibly a storm or upwelling, in the southern hemisphere
at approximately 50$\fdg$S latitude.  The spectrum also includes 
light from a dimmer northern feature at approximately 30$\fdg$N latitude.
We compare our spectra ($\lambda/\Delta\lambda\sim$2000) 
of these two features
with a simple model of Neptune's atmosphere.  Given our model 
assumption that the clouds are flat reflecting layers, we find
that the top of the bright southern cloud feature sat at a pressure level
of 0.14~$^{+0.05}_{-0.03}$~bar, and thus this cloud
did not extend into the stratosphere (P$<\sim$0.1~bar). 
A similar analysis of the dimmer northern feature gives a cloud-top
pressure of 0.084$\pm$0.026 bar.
This suggests that the features we observed efficiently transport
methane to the base of the stratosphere, but do not 
directly transport methane to the upper stratosphere
(P$<10^{-2}-10^{-3}$~bar) where photolysis occurs.
Our observations do not constrain 
how far these clouds penetrate down into the troposphere.
We find that our model fits to the data restrict 
the fraction of H$_{2}$ in \textit{ortho/para} thermodynamic
equilibrium to greater than 0.8.
\end{abstract}




\section{Introduction}

The first hint of Neptune's atmospheric complexity and variability
came when \citet{joyce} observed significant changes in Neptune's 
brightness at $1-4~\mu$m over the course of approximately an Earth
year.  \citet{pilcher} 
interpreted this as the formation and slow dissipation of an
extensive high-altitude cloud.  The 1989 flyby of the Voyager II 
spacecraft revealed a host of time-varying atmospheric features 
\citep{smith1989}.  Even before the Voyager II flyby, the development
of the Charge Coupled Device (CCD) allowed imaging of Neptune at
wavelengths up to $\sim$1~$\mu$m.  Several observers looking in
the 0.62 and 0.89~$\mu$m methane absorption bands regularly found
mid-latitude features that were extremely bright relative to
Neptune's disk  \citep{smith1984,smith1985,H1987,H1989,Hetal1989,H1990}.
These features are presumably clouds and may
be storms or large upwellings of material from the troposphere.
When present, the reflected sunlight from these features dominates
images of Neptune at methane-absorbing wavelengths between 0.6
and 2.5~$\mu$m, as shown by many observers.  
Hubble Space Telescope (HST) regularly observed such
features at wavelengths less than 1~$\mu$m  \citep{S1995,H1995,H1997},
while ground based observers using conventional
infrared techniques have seen these features at 1 to 2.5~$\mu$m 
\citep{S2001a,S2001b,S2001c}.  More recently, high resolution
techniques such as speckle imaging and adaptive optics (AO) have been
used to observe Neptune and these bright features at
1-2.5~$\mu$m  \citep{roddier1997,roddier1998,roe,gibbard,max}.

Speculation about the nature and origin of these phenomena has primarily
focused on the idea of large upwellings punching through the
tropopause resulting in a high column density of condensed methane
particles.  Thus, these features could in part be responsible for 
transporting methane through the cold-trap of the tropopause
and loading the stratosphere with methane gas, 
where it is then photolyzed and converted
to a variety of heavier hydrocarbons, eventually forming hazes 
\citep{B1995b,romani,moses}.
It is crucial for our understanding of the dynamics and chemistry
of Neptune's atmosphere to know the altitude range to which 
these cloud features reach.

\citet{Hetal1989} estimated from their CCD photometry that the bright
features they observed were due to increases in the number density of 
high stratospheric haze particles.
\citet{roddier1998} observed Neptune with adaptive optics techniques.
They used two narrowband filters centered on 1.56 and 1.72~$\mu$m, 
such that one filter is centered on a 
strong methane absorption feature while the other filter is outside
the strong methane absorption.  
These authors estimated that the bright features are located near the
tropopause at pressures on the order of 0.1~bar or, possibly, at even 
higher altitudes.  More recently \citet{S2001c}, using IRTF photometry,
found the altitudes of a number of discrete cloud features to be between
0.060 and 0.230~bar.  In this paper we present spectra of two of these
cloud features.  Through comparison with a simple
radiative transfer model, we used these spectra to 
determine precisely the altitude of the cloud features.
Our best-fit model places the top of the bright southern 
cloud feature that we observed
at a pressure level of 0.14~bar within an uncertainty range of
0.11 to 0.19~bar, while the best-fit for the dimmer northern feature
puts it at 0.084$\pm$0.026 bar.

\section{Observations and Data Reduction}

We observed Neptune on 1999 June 2 (UT) using NIRSPEC, the 
W.M.\ Keck Observatory's new near-infrared spectrometer, on the 
Keck II telescope during the commissioning of the NIRSPEC
instrument \citep{mclean}.  This spectrometer operates over a
wavelength range of 0.95--5.5~$\mu$m in either
a low-resolution (R $\sim$ 2000) mode or a cross-dispersed 
high-resolution (R $\sim$ 25,000) mode.  NIRSPEC is equipped with a 
1024 $\times$ 1024 InSb ALADDIN array for spectroscopy, and
also a slit-viewing camera (SCAM) containing a 256 $\times$ 256
HgCdTe PICNIC array with 0$\farcs$18 pixels. 
The data presented here 
are from a single low-resolution setting using the NIRSPEC-6 
blocking filter, and they cover roughly 2.03--2.30~$\mu$m.   In 
low-resolution mode the pixel size in the spatial direction of
the ALADDIN spectral array is 0.144''/pixel.

We acquired a series of slit-viewing camera (SCAM) images both
before and during our spectral exposures, giving us images of Neptune
with the spectrometer's slit offset from the disk of Neptune and overlapping
the disk of Neptune.  SCAM images were taken in pairs, and after the
first exposure of each pair the pointing of the telescope was
offset by 10''.  In the current work we did not attempt
precise photometry, and therefore our processing of the SCAM
images is simplistic: We subtracted one image of each pair from the
other for background and bias subtraction.  We then shifted and coadded the 
images using the Gaussian centroids of Triton and an
unidentified star for offset determination.  From Triton (apparent
diameter 0$\farcs$126) and the unidentified star the 
FWHM of the point spread function was 0$\farcs$43$\pm$0$\farcs$07.

Figure 1a shows a SCAM image of Neptune taken simultaneously with the
spectra presented in this work where the disk is bisected by the slit.
Figure 1b shows an unobstructed
image taken 25 minutes earlier.  The SCAM image shown in Fig.\ 1a
was taken with the NIRSPEC-6 filter, (1.56-2.32~$\mu$m), 
while the unobstructed SCAM image in Fig.\ 1b was taken with the
NIRSPEC-7 filter (1.84-2.63~$\mu$m).  We did not take an unobstructed
SCAM image in the NIRSPEC-6 filter, and therefore we present the 
NIRSPEC-7 filter image to show Neptune unobstructed by the
spectrometer slit.
Neptune's apparent diameter (at 1~bar level)
was 2$\farcs$30, the Earth's planetographic 
sub-latitude on Neptune was -28$\fdg$08, 
and the solar phase angle was 1$\fdg$54.\footnote{See the NASA/JPL 
Horizons ephemeris program at http://ssd.jpl.nasa.gov/horizons.html.}   
Figure 1c shows the orientation and scale of Neptune on the images in
Fig.\ 1a-b.  Neptune's brightness along
the slit is shown in Fig.\ 1d.  Comparison of Neptune's brightness as a
function of position on the slit with that of
HD201941 (Fig.\ 1d) shows that the projected size of the storm
is only marginally resolved.  The local minimum between the features
on Neptune indicates that we can extract spectra of the two separate
features relatively cleanly without much cross-contamination.

\begin{figure}
\vskip -1.5in
\hskip -0.5in
\epsfig{figure=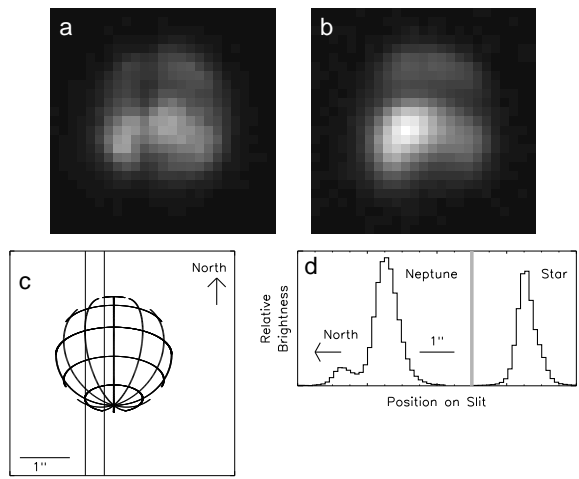,width=4in,angle=0}
\vskip -1in
\caption{\small (a)  SCAM image of Neptune with spectrometer 
slit overlying Neptune.  This image was taken in the NIRSPEC-6 filter 
(1.558-2.31~$\mu$m) at 14:06 UT during the first Neptune spectral
exposure. (b) SCAM image of Neptune unobstructed by slit.  This image was taken
in the NIRSPEC-7 filter (1.839-2.630~$\mu$m) at 13:45 UT.  No unobstructed
images of Neptune were taken in the NIRSPEC-6 filter.
(c) Schematic showing the scale and orientation of Neptune in (a) and
(b).  North is up and Neptune appears as it would look on the sky.
The scale of the schematic is shown at lower-left and is the same
as in (a) and (b).  The vertical lines show the approximate location
of the slit in (a).  (d) Brightness
of Neptune along the slit, averaged over all the wavelengths shown in 
the spectral image in Fig.\ 2b.  
For comparison to atmospheric seeing, the profile of the star
HD201941, averaged over the same wavelength range, is also shown.  The profile
of HD201941 gives a seeing FWHM of $\sim0\farcs5-0\farcs6$.  The FWHM of
the unidentified star in the the scam images taken simultaneously with
the Neptune spectra was 0$\farcs$4.  The narrowness of the FWHM
compared to the separation of Neptune's northern and southern features,
along with the clear local minimum between the features, leads us to 
conclude that we have separated   the light from these
two features relatively well.
\label{fig1}}
\end{figure}

The spectra presented here come from two 60-second exposures taken
with a 42$\farcs$ $\times$ 0$\farcs$380 slit starting at 14:06 (UT) on
1999 June 2.  
The slit was aligned parallel with Neptune's north-south axis
and centered on the bright feature in the southern hemisphere.
This feature was by far the brightest that we observed on Neptune 
on 1999 June 2 (UT).  The slit also captured light from a dimmer
feature in Neptune's northern hemisphere.
Between the two exposures the pointing of the telescope was moved
$\sim$10'' along the direction of the slit, so that Neptune fit easily on
the slit for both exposures.  In order to correct for Earth's
atmospheric absorption, we observed an A2 spectral type   
star (HD201941) in two 10 second exposures, with an offset of the
telescope pointing between exposures in order to move the star along
the slit.  

The reduction sequence consisted simply of subtraction of one star spectral
frame from the other for bias and background subtraction.  
In these images the spatial and spectral
coordinates are distorted with respect to the rows and columns of 
the detector array.  
The OH sky emission lines in unsubtracted frames
trace out lines of constant wavelength and are well fit by straight
lines
\begin{equation}
 Y(x) = A_{0}(\lambda) + A_{1}(\lambda) x,
\end{equation}
where A$_{0}$ and A$_{1}$ are functions of wavelength.
Meanwhile, the arc of a stellar spectrum traces out a line of 
constant position along the slit and is well fit by the function
\begin{equation}
 X(y) = B_{0}(s) + B_{1}(s) y + B_{2}(s) y^{2},
\end{equation}
where B$_{0}$, B$_{1}$, and B$_{2}$ are all functions of slit position
(s).  The (x,y) position on the array for a given slit-position and
wavelength (s,$\lambda$) can then be found by: interpolating A$_{0}$ 
and A$_{1}$ for $\lambda$ from the numerous OH lines that we fit,
interpolating B$_{0}$, B$_{1}$, and B$_{2}$ from the several stellar
spectra fit, and finally finding the (x,y) intersection of Eqns.\ 1
and 2.  By doing this for a grid of wavelengths and positions along the
slit the data in a spectral image is interpolated from (x,y) to (s,$\lambda$).
The final step in this rectification process 
is to apply a Jacobian correction for the geometric distortion.
We extracted the stellar spectrum from the rectified spectral image
using an optimal weighted extraction technique that includes a median 
filter rejection algorithm to remove the effects of bad pixels and
cosmic ray hits.  Finally, we
divided the extracted stellar spectrum by that of Vega (spectral
type A0V) \citep{colina2} to produce an estimate of the 
combined atmospheric and instrumental transfer function, shown in Fig.\ 2a.

\begin{figure}
\vskip -3.0in
\hskip -0.75in
\epsfig{figure=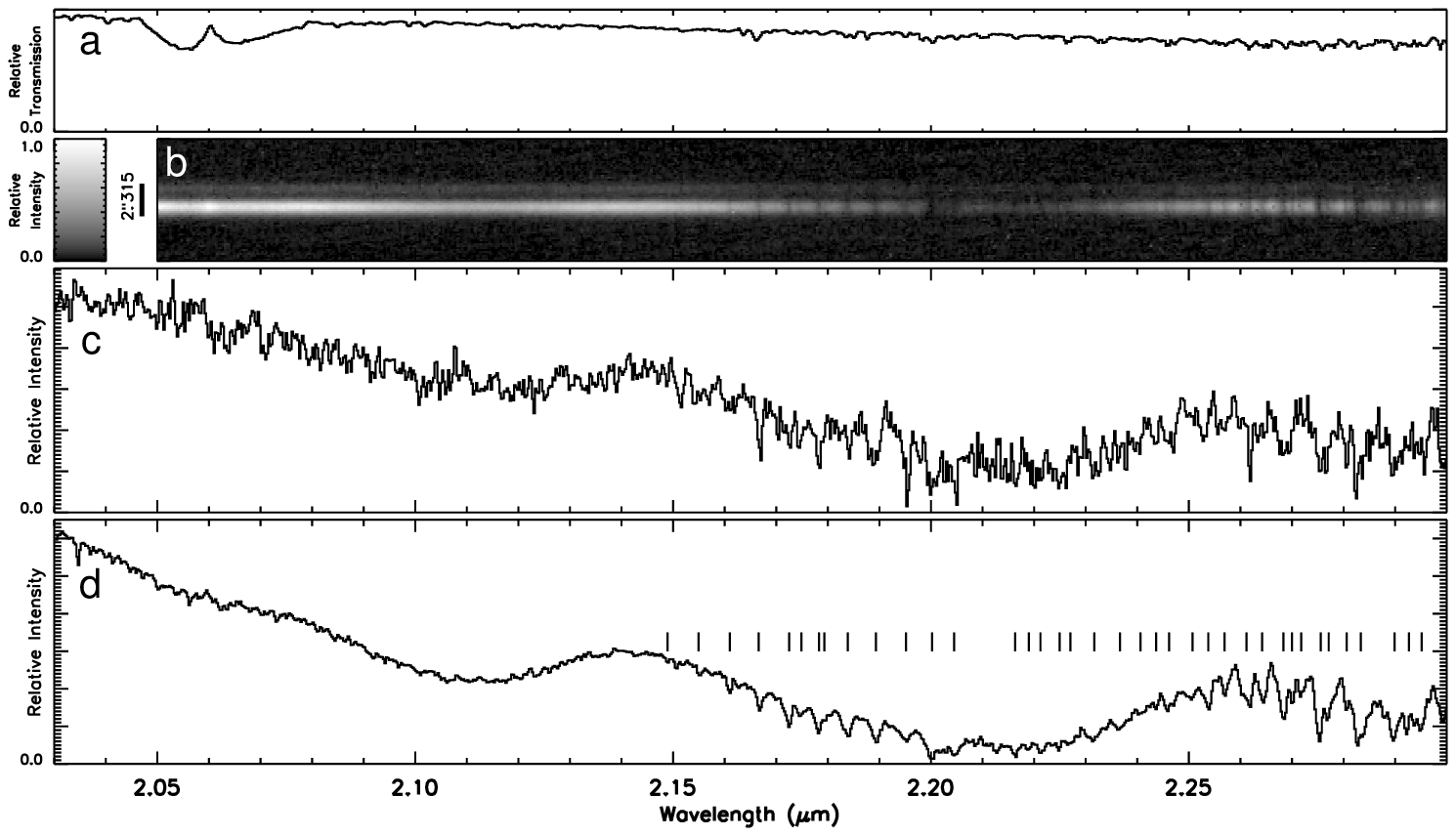,width=4.5in}
\vskip -1.75in
\caption{\small 
(a) Observed spectrum of A2 star HD201941 divided by the reference
spectrum of A0V star Vega, providing an estimate of the atmospheric
and instrumental transmission functions.  (b) Rectified spectral image
of Neptune.  The spectral region of 2.03 to 2.05 $\mu$m is omitted in
order to show an intensity-bar and the spatial size of Neptune's
diameter along the slit.  (c) Extracted spectrum of the dimmer
northern feature. d) Extracted spectrum of the brighter southern
feature.  Note the short vertical tick marks that show methane
absorption lines identified from the laboratory spectrum of
\citet{mckellar}.  The vertical scale in (c) is exaggerated by a
factor of 11 with respect to the vertical scale in (d).\label{fig2}}
\end{figure}

We processed the spectrum of Neptune
in a manner similar to that applied to the calibration star; 
however, after rectification, but before extraction,
we inserted the additional step of dividing by the transfer function 
determined from the star.  The final rectified spectral image of 
Neptune is shown in Fig.\ 2b.  We extracted the spectra of the
cloud features in a similar manner as for the stellar spectrum, except that
we limited the extractions to 0$\farcs$6 along the slit centered on each
feature.
The spectrum of the dimmer northern cloud feature 
is shown in Fig.\ 2c, and the spectrum of the brighter southern feature
is shown in Fig.\ 2d.  Shown in Fig.\ 2c and 2d are averages 
from the two separate exposures.  Having two separate exposures
provides a check on our precision.  The two spectra of the northern
feature extracted from the two exposures appeared identical except
for random noise.  Similarly, the two southern feature spectra were also
very nearly identical.

Navigation on the images of Neptune is difficult because light
from the cloud so dominates over all other features, however the
presence of Triton in the slit-viewing camera images makes this
problem significantly easier.  By centroiding a gaussian on 
Triton and using the offset from Triton to Neptune's
center given by JPL's Horizons ephemeris, we find the center of
Neptune.  Combining this with centroiding a Gaussian on the 
cloud feature we 
estimate the cloud to be located 0$\farcs$65$\pm$0$\farcs$25 from 
the center of Neptune's disk at a Neptune latitude of 
-48$\fdg\pm$6$\fdg$.  Thus, the cloud lay at a viewing angle
$\Theta$ of 34$\fdg\pm$17$\fdg$, where $\Theta$ is the angle between
the normal on Neptune's `surface' and our line of
sight.  By a similar procedure we estimate the observed northern
feature to be at a latitude of 30$\fdg\pm$13$\fdg$ and a viewing
angle of 55$\fdg^{+9\fdg}_{-4\fdg}$.

\section{Atmospheric Model}

Our aim in the work presented here is to measure the altitude
or pressure level at the top of an infrared-bright cloud on Neptune.  
Towards this end we have taken spectra, presented in the previous
section, over a wavelength range where the opacity in Neptune's 
atmosphere varies significantly as a function of wavelength due 
to H$_{2}$ collision induced absorption (H$_{2}$-CIA) and methane
absorption.  In
our model we calculate the predicted spectrum as a function of the
altitude of the top of the cloud and several other parameters
described below.  We judge the goodness-of-fit for each model
spectrum using the metric $\sum ( I_{obs}(\lambda) - A \cdot 
I_{model}(\lambda) )^{2}$, where $A$ is chosen in each case to 
minimize the overall sum.  The introduction of the factor $A$
is necessary due to the lack of an absolute flux calibration for
the observed spectrum, I$_{obs}(\lambda)$.

Our model atmosphere consists of 120 layers evenly spaced
in Log$_{10}$(P$_{bar}$) from 5.0 to $10^{-4}$~bar.  We interpolate
the temperature and pressure for each layer from \citet{lindal}.
The free parameters in our model are: the mole fraction
of helium, F$_{He}$; the mole fraction of methane in
the stratosphere, F$_{CH_{4},s}$; the mole fraction of 
methane in the troposphere, F$_{CH_{4},t}$; the fraction of H$_{2}$
in \textit{ortho/para} thermodynamic equilibrium, F$_{eq}$; the
viewing angle, $\Theta$; and the pressure altitude of the top
of the cloud in~bar, P$_{bar}$.
Around the tropopause the fractional methane abundance follows
the saturation vapor curve, so that the methane
abundance is never super-saturated.  Wavelengths of 2.03-2.30 
$\mu$m do not probe significantly into the troposphere, and therefore
our model fit is insensitive to changes in F$_{CH_{4},t}$.
In each layer a fraction of the H$_{2}$, F$_{eq}$, is distributed 
between \textit{ortho} and \textit{para} states according to
thermodynamic equilibrium, with the remaining H$_{2}$ distributed
according to an \textit{ortho}:\textit{para} ratio of 3:1. 
We calculated the model
predicted spectrum for each point on a grid of these
free parameters.  The grid points for each parameter are listed
in Table 1.

\begin{deluxetable}{cl}
\tabletypesize{\scriptsize}
\tablecaption{Model parameter grid points. \label{tbl-1}}
\tablewidth{0pt}
\tablehead{
\colhead{Parameter} & \colhead{Values used in model}   
}
\startdata
F$_{He}$ $^{a}$  &  0.08, 0.10, 0.12, 0.14, 0.16, 0.18, 0.20, 0.22 \\
F$_{CH_{4},s}$ $^{b}$  &  $2.\times10^{-5}, 8.\times10^{-5}, 
        1.8\times10^{-4}, 3.5\times10^{-4}, 7.0\times10^{-4}, 
        1.05\times10^{-3}, 1.7\times10^{-3}$ \\
F$_{CH_{4},t}$ $^{c}$  &  0.022 \\
F$_{eq}$ $^{d}$  & 0.0, 0.5, 0.6, 0.7, 0.8, 0.9, 1.0  \\
$\Theta$  &   17$\fdg$, 34$\fdg$, 51$\fdg$ \\
P$_{bar}$  &  120 layers evenly spaced in $Log_{10}(P)$ 
              from 5.0 to 10$^{-4}$~bar \\

 \enddata


\tablenotetext{a}{\citet{conrath1993} found F$_{He}=0.15$}.
\tablenotetext{b}{\citet{BH1994} found F$_{CH_{4},s}=3.5\times10^{-4}$ with a
    maximum uncertainty range of $2.5\times10^{-5}$ to $1.7\times10^{-3}$.}
\tablenotetext{c}{\citet{B1995a} found F$_{CH_{4},t}=0.022$.  The fit
of our model to data is not sensitive to F$_{CH_{4},t}$.}
\tablenotetext{d}{\citet{BS1990} found F$_{eq}=1.0$, with a minimum 
allowed value of 0.85.}


\end{deluxetable}

To model collision-induced absorption by hydrogen (H$_{2}$-CIA) for
H$_{2}-$H$_{2}$ and H$_{2}-$He collisions we
use the FORTRAN routines of A. Borysow \citep{borysow1,borysow2,
borysow3,borysow4,borysow5}.\footnote{Available 
at: http://www.astro.ku.dk/$\sim$aborysow/programs/index.html}
Although both 0-1 and 0-2 transitions are included in our model, for
wavelengths of 2.1--2.3~$\mu$m only the 0-1 transition 
is relevant. 

Accurate modeling of methane absorption across the near-infrared 
spectrum is extremely difficult due to the enormous number of 
individual lines and huge variation in line strength.  We 
apply the correlated \textit{k} distribution method as 
described in \citet{lacis} and on p.\ 230 of \citet{goody}.  We use the
H$_{2}$-broadened methane \textit{k}-coefficients of \citet{irwin}
since Neptune's atmosphere is primarily H$_{2}$.  These coefficients
are for 5 cm$^{-1}$ wide bins and this places a limit on the spectral
resolution of the model.  

We ignore all scattering processes, except reflection
from the top of the cloud.  Although significant at shorter 
wavelengths, Rayleigh scattering is negligible at wavelengths of
2.0 to 2.3~$\mu$m.  Light reflected from the top of the cloud
dominates all other sources which might contribute to our cloud
spectrum, such as scattered light from 
stratospheric hydrocarbon hazes.  Therefore, 
the only scattering process that we include is reflection from the top
of the cloud, which we model as a flat reflecting layer.  
The reflectivity of the top of the cloud,
or alternatively, the combined optical depth, scattering phase function,
and single scattering albedo, are irrelevant given that we fit the
shape of the spectrum, not the absolute flux level.  Further,
we assume that the optical depth, scattering phase function,
and single scattering albedo do not vary significantly over the
wavelength range of our spectrum (2.03 to 2.3~$\mu$m).  
Using the solar spectrum of \citet{colina1}, 
the model produces the estimated spectrum of
the cloud as a function of the pressure level of the top of the cloud.

\section{Results of Model Fit}

The spectral resolution of the  model described in the previous 
section is limited by the \textit{k}-coefficients of \citet{irwin}
and is coarser than the spectral resolution of our observed spectra.  
Therefore, we bin the observed spectra in wavelength to
achieve a resolution as nearly identical as possible to the model
spectra.  We restrict the model fitting to the wavelength range
2.08-2.25~$\mu$m in order to avoid a large telluric CO$_{2}$ band
at $<2.08 \mu$m and a series of sharp methane features at
$>2.25 \mu$m that are poorly represented by the methane coefficients
in the model.

For the bright southern feature at 
a viewing angle of $\Theta = 34\fdg$, we obtain our best-fit
for a cloud top at 0.14~bar, F$_{CH_{4},s} = 0.0017$, 
F$_{He} = 0.22$, and F$_{eq} = 1.0$.  For the dimmer northern feature
at a viewing angle of $\Theta = 55\fdg$, the best fit is for a cloud-top
pressure of 0.084~bar, F$_{CH_{4},s} = 0.0017$, 
F$_{He} = 0.22$, and F$_{eq} = 1.0$.  These best-fit model spectra
are superposed on the observed spectra in spectrum (1) of Fig.\ 3a and
spectrum (1) of Fig.\ 3b.  The two parameters that we can best 
constrain are P$_{bar}$ at the top of cloud and the fractional 
equilibrium of H$_{2}$ in \textit{ortho/para} equilibrium, F$_{eq}$.
The model spectra do not fit the observations for 
cloud-top pressures outside the range of 0.11-0.19~bar for the bright
southern feature (see spectra (2) and (3) of Fig.\ 3a), nor outside
the range of 0.058-0.110~bar for the dimmer northern feature 
(see spectra (2) and (3) of Fig.\ 3b).  As shown in Fig.\ 4, reasonable
model fits to both the northern and southern spectra require
F$_{eq}>\sim$0.8.
We find that our data do not constrain significantly 
F$_{He}$ and F$_{CH_{4},s}$.

\begin{figure}
\vskip -1.0in
\hskip -0.75in
\epsfig{figure=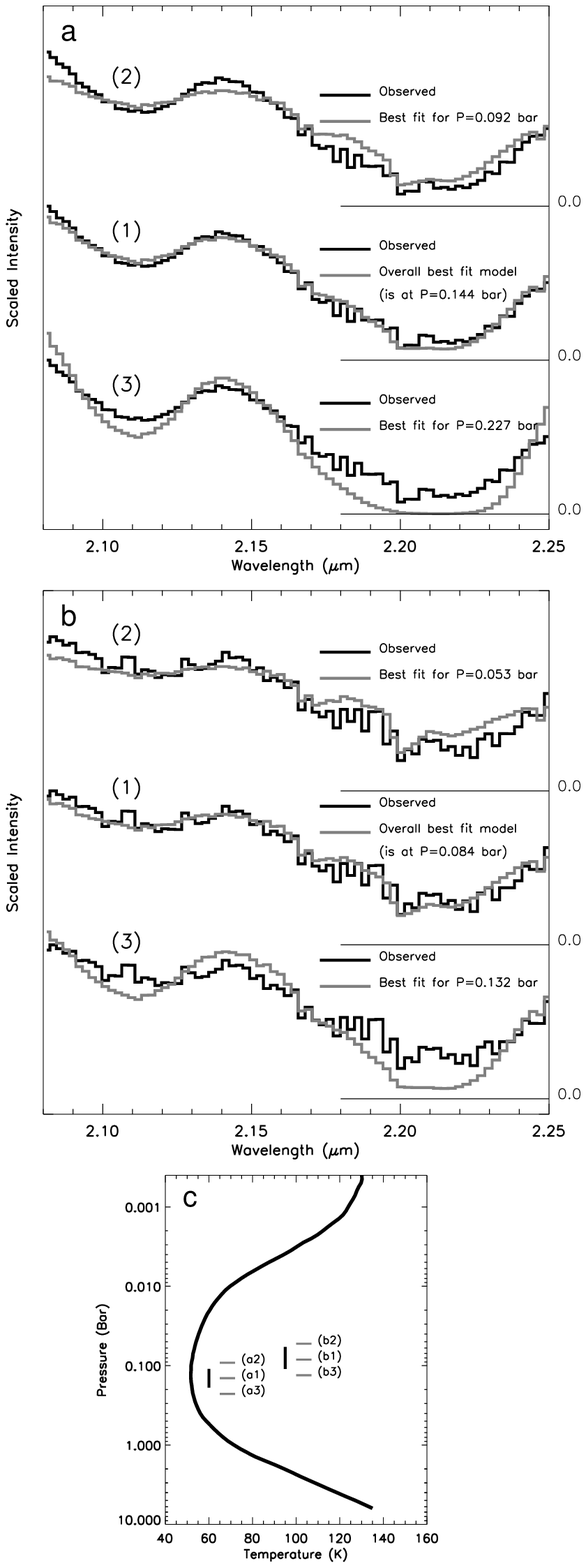,width=4.5in}
\vskip -0.5in
\caption{\small Constraint on the cloud top pressure.
(a) In each case the solid dark line is the binned observed spectrum of
the bright southern feature.  
The overplotted grey lines
are the model spectra:  (1) The overall best fit model
is for P$_{bar}$=0.144~bar, F$_{CH_{4},s}=1.7\times10^{-3}$, F$_{He}$=0.22,
F$_{eq}$=1.0, and $\Theta =34\fdg$.  (2) When the cloud top is raised to
0.092~bar, the best fit is poor and requires F$_{CH_{4},s}=1.7\times10^{-3}$,
F$_{He}=0.08$, F$_{eq}=1.0$, and $\Theta=34\fdg$.
(3)  Similarly, when the cloud top is lowered to 0.227~bar,
the best fit is poor.  In this case F$_{CH_{4},s}=2.\times10^{-5}$, 
F$_{He}=0.22$, F$_{eq}=1.0$, and $\Theta=34\fdg$.  
(b)In each case the solid dark line is the binned observed spectrum of
the dimmer northern feature.  
The overplotted grey lines
are the model spectra:  (1) The overall best fit model
is for P$_{bar}$=0.084~bar, F$_{CH_{4},s}=1.7\times10^{-3}$, F$_{He}$=0.22,
F$_{eq}$=1.0, and $\Theta =55\fdg$.  (2) When the cloud top is raised to
0.053~bar, the best fit is poor and requires F$_{CH_{4},s}=1.7\times10^{-3}$,
F$_{He}=0.08$, F$_{eq}=1.0$, and $\Theta=55\fdg$.
(3)  Similarly, when the cloud top is lowered to 0.132~bar,
the best fit is poor.  In this case F$_{CH_{4},s}=1.05\times10^{-3}$, 
F$_{He}=0.22$, F$_{eq}=0.9$, and $\Theta=55\fdg$.  
(c) The temperature-pressure profile of \citet{lindal}.
Also plotted are the cloud top pressures of the model spectra
in Fig.\ 3a.  The short vertical line on the far
left shows the narrow range of pressures to which our observations
restrict the top of the cloud.  \label{fig3}}
\end{figure}

Due to the low spatial resolution of our data there is
significant uncertainty in the viewing angle for both features, 
$\Theta=34\fdg\pm17\fdg$ for the bright southern feature and 
$\Theta=55\fdg^{+9\fdg}_{-4\fdg}$ for the dimmer northern feature.
Decreasing $\Theta$ to $17\fdg$ for the bright southern feature 
pushes the best-fit cloud-top pressure to 0.16 bar, while 
increasing $\Theta$ to 51$\fdg$ changes the best-fit cloud-top 
pressure to 0.12~bar.  Similarly for the northern feature, 
decreasing $\Theta$ to 51$\fdg$ moves the best-fit cloud-top pressure
to 0.092~bar, while increasing $\Theta$ to 64$\fdg$ shifts the best-fit
cloud-top pressure to 0.076~bar.  In all these cases the best-fit
parameters include F$_{eq}=1.0$ and F$_{CH_{4},s}=0.0017$.

\section{Errors and Uncertainties}

At this point it is worthwhile to make a brief discussion of how 
the errors and uncertainties 
in our observations and model fitting could affect our results
with respect to cloud-top pressure and F$_{eq}$.  

On the observing side, we are much more concerned with systematic
errors, for instance artificial slopes across the entire spectrum,
than with random errors in the spectra of HD201941 and
Neptune.  Since we are fitting the model to 73 wavelength bins,
random errors from bin to bin will tend to cancel out and not bias
the model fit.  There are several possible sources of systematic
errors on the observing side; the three of greatest concern relate to
alignment on the slit and the method of atmospheric correction.  
Misalignment of the slit on the star would redden the spectrum and possibly 
introduce a bias in the final model fitting.  This is less
of an issue on an extended source such as the clouds on Neptune.
By looking at multiple stellar spectra we estimate that this source
of error introduces at most a one to two percent slope from 2.08 to
2.25~$\mu$m.

\begin{figure}
\vskip -2.0in
\hskip -1.5in
\epsfig{figure=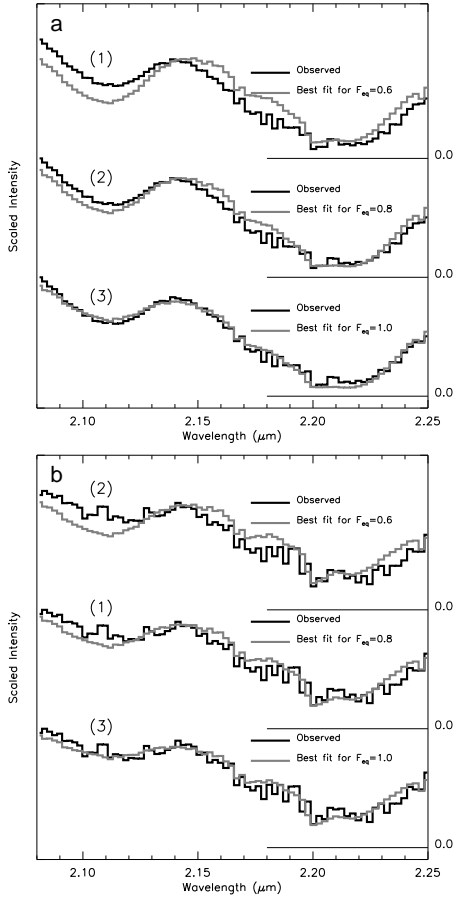,width=6in}
\vskip -1.5in
\caption{\small Constraint on the fraction of H$_{2}$ in \textit{ortho/para} 
equilibrium (F$_{eq}$).
(a)  In each case the solid dark line is the binned observed spectrum of
the bright southern feature.  Holding F$_{eq}$ fixed at 0.6, 0.8, 1.0,
the best fits of model to data are shown.  In all three cases the best fit
included P$_{bar}$=0.144~bar and F$_{CH_{4},s}1.7\times10^{-3}$.  In
cases (1) and (3) F$_{He}$=0.22, while in case (2) F$_{He}$=0.20, which
is not a significant difference.
(b)  In each case the solid dark line is the binned observed spectrum of
the dimmer northern feature.  Holding F$_{eq}$ fixed at 0.6, 0.8, 1.0,
the best fits of model to data are shown.  In all three cases the best fit
was P$_{bar}$=0.144~bar, F$_{CH_{4},s}1.7\times10^{-3}$, and F$_{He}$=0.22.
\label{fig4}}

\end{figure}

In applying the atmospheric and instrumental transmission 
correction with HD201941 there are
two more potential sources of systematic error.  
The first is that to find the atmospheric transmission function
we divided HD201941 by a spectrum of Vega, and the second is that
HD201941 was not observed at exactly the same airmass and time as Neptune.  
While Vega is an A0V star,
HD201941 is listed as an A2 star in the SIMBAD database.\footnote{Available at
http://simbad.u-strasbg.fr.}  In order to estimate the maximum slope
bias that this stellar mis-match could introduce, we compared blackbody
curves.  \citet{drilling} give the T$_{eff}$ for an A0V star as 9790~K
and for an A2V star as 9000~K.  This difference suggests a slope error
of 0.29 percent from 2.08 to 2.25~$\mu$m.  The spectra of HD201941 were
taken at an airmass 1.05, while the Neptune spectra were taken at airmass
1.28.  In order to minimize the influence of this on the model fitting
we excluded wavelengths shortward of 2.08~$\mu$m to avoid a large CO
band.  To investigate what biases and slopes this mismatch in airmass
could introduce we used the ATRAN \citep{lord}
model atmospheric transparency spectra
available on the Gemini Observatory website.\footnote{See 
http://www.gemini.edu/sciops/telescope/telIndex.html.}  While
we did not examine transparency spectra for our exact airmasses, the 
slope difference introduced by observing at airmass 1.0 versus 1.5 
across our spectral range of interest would be 1.6 percent.

Each of these possible slope errors discussed above is less than
2 percent.  To show that even a fortuitous addition of all these
slope errors in one direction would not change our primary results
we artificially introduced slope errors of $\pm$10 percent to our 
final observed spectra and refit the model.  This had no effect
on results concerning F$_{eq}$ and at most shifted the best-fit
pressure level of the cloud top by one level in our model, to 0.12~bar
in the $-10\%$ case for the bright southern feature and to 0.09~bar 
in the $+10\%$ case for the dimmer northern feature. 

While there are numerous small ways in which the model may be inaccurate,
for instance if the temperature-pressure curve is not exactly correct
for the location of the cloud features, the two major sources of
uncertainty in the model are the methane \textit{k}-coefficients
of \citet{irwin} and the assumption of a flat reflecting cloud layer.
For Neptune's atmosphere we are forced to extrapolate the methane 
\textit{k}-coefficients to much colder temperatures than the temperatures
of the laboratory measurements on which they are based.  While the 
accuracy or inaccuracy of this extrapolation is difficult to judge, 
the independence of best-fit cloud-top pressure and F$_{eq}$
from methane concentration F$_{CH_{4},s}$ gives us confidence in our
results.

For ease we assume in our model that the cloud top is a flat
reflecting layer, however due to particle scattering properties the
reflectivity of the cloud may vary with wavelength and the `top' of the
cloud is almost certainly somewhat extended.  Our best-fit model for
the bright southern feature (see spectrum 1 of Fig.\ 3a) is systematically
slightly off from the observed spectrum.  This is most easily seen
at wavelengths shortward of $\sim$2.16~$\mu$m where H$_2$ absorption
dominates.  This is suggestive that while a flat reflecting layer
at 0.144~bar does not fit perfectly, a combination of reflectance
from pressures slightly higher to slightly lower than 0.144~bar might
result in a better fit, which is exactly what one would expect if
the cloud-top were somewhat extended.  In fact, a more detailed method
of modeling would be to view the computed model spectra for all the
pressure levels in the model as a basis set and to construct the best-fit
model spectrum from a linear combination of all the spectra from
different levels.  However, one shortfall to that approach is that it
implicitly assumes single scattering, ignoring multiple scattering
between layers.  A more complete modeling approach must include 
the multiple scattering between layers as well, which is beyond the scope
of the current paper.  Variation in reflectivity
as a function of wavelength may also play a role in these slight
discrepancies between model and data.

\section{Conclusions}

By comparing a near-infrared spectrum with the predictions of a simple
transmission model we determined the pressure level at the top
of an infrared-bright tropospheric cloud on Neptune.  We find a 
best-fit of model to data for a cloud-top pressure level of 0.14~bar
within a maximum allowed range of 0.11 to 0.19~bar for the bright
southern feature that we observed.  We found the dimmer northern feature 
to sit slightly higher in the atmosphere at 0.084~bar within a maximum
allowed range of 0.058 to 0.11~bar.
  Our work places no limit on the pressure at the bottom of
the cloud.  Our results further restrict the fraction of H$_{2}$ in
\textit{ortho/para} equilibrium to greater than 0.8, and our best-fits
consistently put this fraction at 1.0.  This is in agreement with
the work of \citet{BS1990} who found the same results, but from a 
different technique, 
measuring the equivalent widths of the 4-0 S(0) and S(1) transitions 
between 0.6 and 0.7~$\mu$m.  Our results
do not constrain the fractional abundance of methane in the
stratosphere, nor the fractional abundance of helium.  

Our primary result is the tight constraint we place on 
the pressure at the top of the
cloud.  By constraining the cloud-top to pressures around the tropopause, we
show that the cloud, possibly a storm or upwelling, does not extend
significantly into the stratosphere.  
If the cloud is made up of condensed methane particles
brought up from below, then the mechanism by which this cloud was formed
appears to be efficient at bringing methane to near the top of the 
troposphere, but, at least at the time we observed, the mechanism
was not acting as
an efficient method of transporting methane to the upper stratospheric
levels where ultraviolet photolysis occurs (P$>10^{-2}-10^{-3}$~bar).

In the current paper we present a measurement of the altitude of 
two cloud features at a single time.
We expect longer-term observations of multiple infrared-bright
features will find that most reach only to approximately the
tropopause, as in the case presented here, but occasional features
may reach far into the stratosphere (P$> 0.01$~bar) and thus would 
provide an extremely efficient method of transporting methane to
the upper stratosphere for photolysis.  
We are currently undertaking such a program of observations using
NIRSPEC coupled to the Keck Adaptive Optics system \citep{wiz}
to achieve simultaneous high-spatial and high-spectral resolution.

\acknowledgements

H.G.R.\ acknowledges support from a NASA GSRP grant funded through
NASA Ames Research Center and a
Sigma Xi Grant-in-Aid-of-Research from the National Academy of Sciences,
through Sigma Xi, The Scientific Research Society.
This work was
partially supported by the Department of Energy under contract 
W-405-ENG-48 to the University of California Lawrence Livermore National
Laboratory.





\clearpage



\end{document}